\documentclass[prd,reprint,showpacs,showkeys]{revtex4}
\usepackage{eurosym}
\usepackage{amsfonts}
\usepackage{mdframed}
\usepackage{amssymb}
\usepackage{amsmath}
\usepackage{graphicx}
\usepackage[font={footnotesize,it}]{caption}

\begin{document}

\title{Particle Collision near 1+1- Dimensional Horava-Lifshitz Black Hole and Naked Singularity}
\author{M. Halilsoy}
\email{mustafa.halilsoy@emu.edu.tr}
\author{A. Ovgun}
\email{ali.ovgun@emu.edu.tr}
\affiliation{Physics Department, Eastern Mediterranean University, Gazimagusa, TRNC, Mersin 10, Turkey.}
\date{\today}

\begin{abstract}
The unbounded center of mass (CM) energy of oppositely moving colliding particles near horizon emerge also in 1+1- dimensional Horava-Lifshitz gravity. This theory has imprints of renormalizable quantum gravity characteristics in accordance with the method of simple power counting. Surprisingly the result obtained is not valid for a 1- dimensional Compton- like process between an outgoing photon and an infalling massless/ massive particle. It is possible to achieve unbounded CM energy due to collision between infalling photons and particles. The source of outgoing particles may be attributed to an explosive process just outside the horizon for a black hole and the naturally repulsive character for the case of a naked singularity. It is found that absence of angular momenta in 1+1- dimensions does not yield unbounded energy for collisions in the vicinity of naked singularities.
\end{abstract}

\maketitle
\section{INTRODUCTION}

It is known that in spacetime dimensions less than four gravity has no life
of its own unless supplemented by external sources. With that addition we
can have lower dimensional gravity and we can talk of black holes,
wormholes, geodesics, lensing effect etc. in analogy with the higher
dimensions. One effect that attracted much interest in recent times is the
process of particle collisions near the horizon of black holes due to
Banados, Silk and West \cite{BSW} which came to be known as the BSW effect.
This problem arose as a result of imitating the rather expensive venture of
high energy particle collisions in laboratory. From curiosity the natural
question arises: is there a natural laboratory ( a particle accelerator) in
our cosmos that we may extract information/ energy in a cheaper way? This
automatically drew attentions to the strong gravity regions such as near
horizon of black holes. Rotating black holes host greater energy reservoir
due to their angular momenta and attentions naturally focused therein first 
\cite{Sotiriou,Lake}. In case the metric is static and diagonal, there are reasons to consider the collision process in the vicinity of a naked singularity as well. 

We note from physical grounds that outgoing particles from the event horizon of a black hole can not occur. Hawking radiation particles/ photons emerge too weak to compare with infalling particles. Thus collision of two particles can only be argued if both are infalling toward the horizon of a black hole. Such a process, however, yield no BSW effect in the non- rotating metrics, which is our main interest in this study. In order to have an unbounded CM energy in a collision process both particles must be taken in the same coordinate frame and in opposite directions. This is possible in the vicinity of a naked singularity whose repulsive effect compells particles/ photons to make collisions with an infalling particle/ photon. From the outset we state that such a collision taking place near the naked singularity in the absence of angular momenta does not yield an unbounded CM energy. To extend our study to cover also collisions near black holes we assume that some unspecified process, such as disintegration decay process of some particles, yield outgoing particles photons while the partners fall into the hole. For a thorough analysis of all these problems covering the ergosphere region of a Kerr black hole, Penrose process, particle collisions etc. one must consult \cite{piran}.

In general one considers the
radial geodesics and upon energy-momentum conservation in the center-of-mass
(CM) frame the near horizon limit is checked whether the energy is
bounded/unbounded. Our aim in this study is to consider black hole solutions
in 1+1- dimensional Ho\v{r}ava-Lifshitz (HL) gravity \cite{costa} and check
the BSW effect in such reduced dimensional theory. Let us remark that at the
Planck scale in higher dimensions the spherical part $r^{2}d\Omega _{D-2}^{2}
$ of the line element is less effective compared to the time and radial
components. For this reason 1+1- dimensions becomes significant at the
Planck scale. For a number of reasons HL gravity is promising as a candidate
for a re-normalizable quantum gravity physics which has been yearned for a
long time \cite{Horava}. The key idea in HL- gravity is the inhomogeneous
scaling properties of time and space coordinates which violates the Lorentz
invariance. Arnowitt-Deser-Misner (ADM) splitting of space and time\cite{ADM}
constitutes its geometrical background. BSW effect in lower/higher-
dimensions has been worked out by many authors \cite%
{Berti,Banados,Jacobson,Zaslavskii,Wei,Zaslavskii2,Saadat,worm,amir,anton,frolov,shoom,pour,pour2,liu,patil,baus,joshi,schh,josji3,malafarina,grib,sharif,jamil1,jamil,ghoshh,sade1,grib1,grib2,ding,yang,harada,zhong,jing,jiang,nemoto,yang2}%
. Following the similar idea we consider black hole solutions and naked singularities in 1+1-
dimensions and search for the same effect in this lower dimension. It should be
added that with 1+1- dimensional HL theory the simplest non-trivial solution
is the solution describing an accelerated particle in the flat space of
Rindler frame. This justifies also the meaning of the vector field $(a_{i})$
as the acceleration in the HL- gravity. The role of Rindler acceleration in
3+1- dimensions as a possible source of flat rotation curves and geodesics
motion has been discussed recently \cite{halilsoy} . It is our belief that
the results in lower dimensions \ are informative for higher dimensions and
as a toy model can play the role as precursors in this regard. Even a
Compton- like process can be considered at the toy level between a massless photon outgoing from a naked singularity and a particle falling into the naked singularity. The diverging CM energy results in the case of
photon-particle collision in 1+1- dimensions under specific conditions.

Organization of the paper as follows. In Section II, we review in brief the
1+1-D HL theory with a large class of black hole and naked singularity solutions. CM energy of
colliding particles near horizon and naked singularity is considered in Section III. Section IV
proceeds with applications to particular examples. The case of
particle-photon collision is studied separately in Section V. The paper ends
with our conclusion in Section VI.

\section{1+1-D HL BLACK HOLE/ Naked Singularity}

HL formalism in 3+1-D makes use of the ADM splitting of time and space
components as follows 
\begin{equation}
ds^{2}=-N^{2}dt^{2}+g_{ij}\left( dx^{i}+N^{i}dt\right) \left(
dx^{j}+N^{j}dt\right)  \label{sol-bh-0}
\end{equation}

where $N(t)$ and $N^{i}$ are the lapse and shift functions, respectively.
The action of this theory is

\begin{equation}
S=\frac{M_{Pl}^{2}}{2}\int d^{3}xdt\sqrt{g}\left( K_{ij}K^{ij}+\lambda
K^{2}+V(\phi )\right)
\end{equation}

where $K_{ij}$ is the extrinsic curvature tensor with trace $K$ and Planck
mass $M_{Pl}.$ $V(\phi )$ stands for the potential function of a scalar
field $\phi $, and $\lambda $ is a constant $(\lambda >1).$ Reduction from
3+1- D to 1+1- D results in the action \cite{costa}

\begin{equation}
S=\int dtdx\left( -\frac{1}{2}\eta N^{2}a_{1}^{2}+\alpha N^{2}\phi ^{\prime
2}-V(\phi )\right)  \label{action2}
\end{equation}

where $\eta $=constant and $\alpha $=constant  will be chosen to be unity
and $a_{1}$=($\ln N)' $ . Let us comment that a `prime' denotes $\frac{d}{dx}.$
We note that also the first term in $S$ is inherited from the geometric part
of the action while the other two terms are from the scalar field source.
For simplicity we have set also $M_{Pl}=1$ .

It has been shown in \cite{costa} that by variational principle a general
class of solutions is obtained as follows

\begin{equation}
N(x)^{2}=2C_{2}+\frac{A}{\eta }x^{2}-2C_{1}x+\frac{B}{\eta x}+\frac{C}{3\eta
x^{2}}  \label{sol-gen-bh}
\end{equation}

in which $C_{2},$ $A,$ $C_{1},$ $B$ and $C$ are integration constants. Ref.%
\cite{costa} must be consulted for the physical content of these constants.

The line element is 
\begin{equation}
ds^{2}=-N(x)^{2}dt^{2}+\frac{dx^{2}}{N(x)^{2}}
\end{equation}

with the scalar field 
\begin{equation}
\phi (x)=\ln \sqrt{2C_{2}+\frac{A}{\eta }x^{2}-2C_{1}x+\frac{B}{\eta x}+%
\frac{C}{3\eta x^{2}}}.  \label{sol-gen-phi}
\end{equation}

Note that the associated potential is 
\begin{equation}
V(\phi (x))=A+\frac{B}{x^{3}}+\frac{C}{x^{4}}.
\end{equation}

and the Ricci scalar is calculated as

\begin{equation}
R=-\frac{2}{\eta }\left( A+\frac{B}{x^{3}}+\frac{C}{x^{4}}\right) .
\end{equation}
There is naked singularity when $A=C_{1}=0$  
and $C_{2}=B=C=\eta=1$, so that there is no horizon for

\begin{equation}
N(x)^{2}=2+\frac{1}{x}+\frac{1}{3x^{2}} .\label{f}
\end{equation}
Another black hole solution reported by Bazeia et. al. \cite{costa} is found by taking  $C_{1}\neq 0$, $C_{2}\neq 0$, $B\neq 0$ and $A=C=0$

\begin{equation}
N(x)^{2}=2C_{2}-2C_{1}x+\frac{B}{\eta x}.
\end{equation}%
This solution develops the following horizons 
\begin{equation}
x_{h}^{\pm }=\frac{C_{2}}{2C_{1}}\pm \sqrt{\Delta },\text{ }\Delta =\frac{%
C_{2}^{2}}{4C_{1}^{2}}+\frac{B}{2\eta C_{1}}.
\end{equation}%
As $\Delta =0$ they degenerate, i.e., $x_{h}^{+}=x_{h}^{-}$.

The Hawking temperature is given in terms of the outer ($x_{h}^{+}$) horizon
as follows 
\begin{equation}
T_{H}=\left. \frac{\left( N(x)^{2}\right) \prime }{4\pi }\right\vert
_{x=x_{h}^{+}}.
\end{equation}%
For the special case $C_{2}=0$, $C_{1}=-M$ and $B=-2M$ the horizons are
independent of the mass $M$: 
\begin{equation}
x_{h}^{\pm }=\pm \frac{1}{\sqrt{\eta }}\;\;(\eta >0)
\end{equation}%
The temperature is then given simply by 
\begin{equation}
T_{H}=\frac{M}{\pi }.
\end{equation}%
This is a typical relation between the Hawking temperature and the mass of
black holes in $1+1$ dimensions \cite{hawking}.

In the case of $C_{2}=1/2$, $B=-2M$, $\eta =1$ and $A=C=C_{1}=0$ \ , it
gives a Schwarzschild-like solution;%
\begin{equation}
N(x)^{2}=1-\frac{2M}{x} \label{Sch}
\end{equation}

On the other hand, the choice of the parameters, for $C_{2}=1/2$,$B=-2M$, $%
C=3Q^{2}$, $\eta =1$ and $A=C_{1}=0$ gives \ a Reissner--Nordstr\"{o}m-like
solution. \ 
\begin{equation}
N(x)^{2}=1-\frac{2M}{x}+\frac{Q^{2}}{x^{2}} \label{RN}
\end{equation}

As in the general relativity we can make particular choice of the parameters so that we end up with a naked singularity instead of a black hole. The choice $Q^2>M^2$ in eq.(\ref{RN}), for instance,  transforms the HL- black hole into a naked singularity at $x=0$. Similarly $M<0$, turns eq.(\ref{Sch}) into a naked singular metric at $x=0$.

\section{CM ENERGY OF PARTICLE COLLISION NEAR THE HORIZON OF THE 1+1 -D HL
BLACK HOLE}

Here we will derive the equations of motion of an uncharged massive test
particle by using the method of geodesic Lagrangian. Such equations can be
derived from the Lagrangian equation, 
\begin{equation}
{\mathcal{L}}=\frac{1}{2}[-N(x)^{2}\left( \frac{dt}{d\tau }\right) ^{2}+%
\frac{1}{N(x)^{2}}\left( \frac{dx}{d\tau }\right) ^{2}]  \label{lag}
\end{equation}%
in which, $\tau $ is the proper time for time-like geodesics ( or massive
particles). The canonical momenta are 
\begin{equation}
p_{t}=\frac{d{\mathcal{L}}}{d\dot{t}}=-N(x)^{2}\dot{t}  \label{pt}
\end{equation}%
and 
\begin{equation}
p_{x}=\frac{d{\mathcal{L}}}{d\dot{x}}=\frac{\dot{x}}{N(x)^{2}}
\end{equation}

The 1+1- D HL black hole have only one Killing vector $\partial _{t}$ . The
associated conserved quantity will be labeled by $E$ . From eq.$(\ref{pt})$, 
$E$ is related to $N(x)^{2}$ as, 
\begin{equation}
-N(x)^{2}\dot{t}=-E  \label{pt2}
\end{equation}

Hence, 
\begin{equation}
\dot{t}=\frac{E}{N(x)^{2}}  \label{pt3}
\end{equation}

The two-velocity of the particles are given by $u^{\mu }=\frac{dx^{\mu }}{%
d\tau }$. We have already obtained $u^{t}$ in the above derivation. To find $%
u^{x}=\dot{x}$, the normalization condition for time-like particles, $u^{\mu
}u_{\mu }=-1$ \cite{BSW,Misner:1974qy} can be used as, 
\begin{equation}
g_{tt}(u^{t})^{2}+g_{xx}(u^{x})^{2}=-1  \label{nor}
\end{equation}%
By substituting $u^{t}$ to eq.$(\ref{nor})$, one obtain $u^{x}$ as, 
\begin{equation}
(u^{x})^{2}=E^{2}-N(x)^{2}  \label{rdot}
\end{equation}%
for which an effective potential $V_{eff}$ can be defined by 
\begin{equation}
(u^{x})^{2}+V_{eff}=E^{2}.  \label{pot1}
\end{equation}%
Now, the two-velocities can be written as, 
\begin{equation}
u^{t}=\dot{t}=\frac{E}{N(x)^{2}}
\end{equation}%
and 
\begin{equation}
u^{x}=\dot{x}=\sqrt{E^{2}-N(x)^{2}}.
\end{equation}

We proceed now to present the CM energy of two particles with two-velocities $%
u_{1}^{\mu }$ and $u_{2}^{\mu }$. We will assume that both have rest mass $%
m_{0}=1$. The CM energy is given by, 
\begin{equation}
E_{cm}=\sqrt{2}\sqrt{(1-g_{\mu \nu }u_{1}^{\mu }u_{2}^{\nu })}
\end{equation}

so 
\begin{equation}
\frac{E_{cm}^{2}}{2}=1+\frac{E_{1}E_{2}}{N(x)^{2}}-\frac{\kappa \sqrt{%
E_{1}^{2}-N(x)^{2}}\sqrt{E_{2}^{2}-N(x)^{2}}}{N(x)^{2}}  \label{ee}
\end{equation}

where $\kappa =\pm 1$ corresponds to particles moving in the same/ opposite
direction with respect to each other. We wish to stress that our concern is
for the case $\kappa =\pm 1$ since no physical particle is ejected from the
black hole. Note that $E_{1}$ and $E_{2}$ are the energy constants
corresponding to each particle. In the case the second term under the square
root is too small than the first one,%
\begin{equation}
\sqrt{E^{2}-N(x)^{2}}\approx \left( E-\frac{N(x)^{2}}{2E^{2}}+....\right)
\end{equation}

so that the higher order terms can be neglected and CM energy of two
particles can be written as \cite{patil}%
\begin{equation}
\frac{E_{cm}^{2}}{2}\approx 1+(1-\kappa )\frac{E_{1}E_{2}}{N(x)^{2}}+\frac{\kappa}{2}
\left( \frac{E_{2}}{E_{1}}+\frac{E_{1}}{E_{2}}\right) \label{CM}
\end{equation}

Obviously the case with $\kappa =+1$, in which the CM energy becomes%
\begin{equation}
\frac{E_{cm}^{2}}{2}\approx 1+\frac{\left( E_{2}^{2}+E_{1}^{2}\right) }{%
2E_{1}E_{2}},  \label{eqn2}
\end{equation}

where the CM energy is independent from metric function, and it gives always a finite energy. On the other hand $\kappa =-1$ gives%
\begin{equation}
\frac{E_{cm}^{2}}{2}\approx 1+\frac{2E_{1}E_{2}}{N(x)^{2}}-\frac{\left(
E_{2}^{2}+E_{1}^{2}\right) }{2E_{1}E_{2}}  \label{eqn1}
\end{equation}

in which it gives unbounded CM energy near the horizon of the HL black
holes provided an outgoing particle mechanism from the horizon is established. Otherwise the yield of two ingoing particles collision remains finite.

\section{SOME EXAMPLES}

\subsection{Schwarzchild-like Solution}

In the case of $C_{2}=1/2$, $B=-2M$, $\eta =1$ and $A=C=C_{1}=0$ \ , it
gives Schwarzschild-like solution where%
\begin{equation}
V(\phi (x))=-\frac{2M}{x^{3}}
\end{equation}%
and%
\begin{equation}
N(x)^{2}=1-\frac{2M}{x}
\end{equation}

For the CM energy on the horizon, we have to compute the limiting value of
eq.$(\ref{ee})$ as \ $x\longrightarrow x_{h}=2M$ , where is the horizon of
the black hole. \ Setting $\kappa =-1$ as is, the CM energy near \ the event
horizon for 1+1 D Schwarzchild BH is

\begin{equation}
E_{cm}^{2}(x\longrightarrow x_{h})=\infty
\end{equation}

 This result for 4-D Schwarzchild Black hole is already
calculated by Baushev \cite{baus}. Hence, the condition of $\kappa =-1,$
when the location of particle 1 approaches the horizon , on the other hand
the particle 2 runs outward from the horizon due to some unspecified physical process yet yields $%
E_{cm}^{2}\longrightarrow \infty $ so there is BSW effect for 1+1
Schwarzchild-like Solution when the condition $\kappa =-1$ is satisfied.

\subsection{ Reissner-Nordstrom-like solution}

On the other hand, the choice of the parameters, for $C_{2}=1/2$,$B=-2M$, $%
C=3Q^{2}$, $\eta =1$ and $A=C_{1}=0$ gives \ the Reissner--Nordstr\"{o}%
m-like solution. \ 
\begin{equation}
N(x)^{2}=1-\frac{2M}{x}+\frac{Q^{2}}{x^{2}}  \label{31}
\end{equation}

and 
\begin{equation}
V(\phi (x))=-\frac{2M}{x^{3}}+\frac{3Q^{2}}{x^{4}}
\end{equation}

so the CM energy \ is calculated by using the limiting value of eq.(\ref%
{eqn1})%
\begin{equation}
E_{cm}^{2}(x\longrightarrow x_{h=M+\sqrt{(M^{2}-Q^{2})}})=\infty
\end{equation}

so there is a BSW effect.

\subsection{The Extremal case of the Reissner-Nordstrom like black hole}

For the extremal case we have with $M=Q$, from eq.(\ref{31})%
\begin{equation}
N(x)^{2}=\left( 1-\frac{M}{x}\right) ^{2}
\end{equation}

so that it also gives the same answer from eq.(\ref{eqn1}) as\ 
\begin{equation}
E_{cm}^{2}(x\longrightarrow x_{h})=\infty .
\end{equation}

\subsection{Specific New Black Hole Case}

The new 3-parametric black hole solution given by Bazeia, Brito and Costa 
\cite{costa} is chosen as%
\begin{equation}
N(x)^{2}=2C_{2}-2C_{1}x+\frac{B}{\eta x}  \label{nw}
\end{equation}%
with the potential%
\begin{equation}
V(\phi (x))=\frac{B}{x^{3}}.
\end{equation}

For the special case $C_{2}=0$, $C_{1}=-M$ and $B=-2M$ we have

\begin{equation}
N(x)^{2}=2Mx-\frac{2M}{\eta x}
\end{equation}%
\ with suitable potential which is%
\begin{equation}
V(\phi (x))=-\frac{2M}{x^{3}}.
\end{equation}

The CM energy of two colliding particles is calculated by taking the
limiting values of eq.(\ref{eqn1})%
\begin{equation}
E_{cm}^{2}(x\longrightarrow x_{h})=\infty
\end{equation}

Hence the BSW effect \ arises here as well.

\subsection{Near Horizon Coordinates}

We have explored the region near the horizon by replacing r by a coordinate $%
\rho $. The proper distance from the horizon $\rho$ \cite{Susskind:2005js}:
\begin{equation}
\rho =\int \sqrt{g_{xx}(x\prime )}dx\prime =\int_{x_{h}}^{x}\frac{1}{%
N(x^{\prime })}dx\prime
\end{equation}

The first example is the Schwarzchild-like solution which is 
\begin{equation}
N(x)^{2}=1-\frac{2M}{x}
\end{equation}

so that proper distance is calculated as 
\begin{eqnarray}
\rho &=&\int_{x_{h}}^{x}(1-\frac{2M}{x})^{-\frac{1}{2}}dx\prime \\
&=&\sqrt{x(x-2M)}+2MG\sinh ^{-1}(\sqrt{x/2M-1}).  \notag
\end{eqnarray}

The new metric is 
\begin{equation}
ds^{2}=-(1-\frac{2M}{x(\tilde{\rho})})dt^{2}+d\tilde{\rho}^{2}
\end{equation}

where $\tilde{\rho}\simeq 2\sqrt{2M(x-2M)}$so that it gives approximately%
\begin{equation}
ds^{2}\simeq -\frac{\rho ^{2}}{\left( 4M\right) ^{2}}dt^{2}+d\rho ^{2}
\end{equation}

which is \ once more the Rindler type line element. Let us note that this Rindler- type line element is valid within the near horizon limit approximation. For practical purposes there are advantages in adapting such an approximation which is also in conform with the equivalence principle \cite{Susskind:2005js}. The CM energy of two
colliding particles is given by%
\begin{equation}
\frac{E_{cm}^{2}}{2m_{0}^{2}}=1+\frac{\left( 4M\right) ^{2}\left(
E_{1}E_{2}-\kappa \sqrt{E_{1}^{2}-\frac{\rho ^{4}}{\left( 4M\right) ^{4}}}%
\sqrt{E_{2}^{2}-\frac{\rho ^{4}}{\left( 4M\right) ^{4}}}\right) }{\rho ^{2}}
\end{equation}

so that there is BSW effect for $\kappa =-1$ when $\rho \longrightarrow 0.$

\section{Particle Collision near the Naked Singularity }

There is a naked singularity for our 1+1- HL model at the location
of $x=0$ , with $Q^2>M^2$ in eq.(\ref{RN}). In addition $M<0$, turns eq.(\ref{Sch}) into a naked singular metric at $x=0$.
There is also naked singularity when we choose metric function as follows
\begin{equation}
N(x)^{2}=2+\frac{1}{x}+\frac{1}{3x^{2}}=\frac{6x^2+3x+1}{3x^2} .\label{f}
\end{equation}

As it is given in eq.(\ref{ee}), CM energy of the collision of two particles generally 
is (for $N(x) \longrightarrow \infty) $
\begin{equation}
\frac{E_{cm}^{2}}{2}\approx 1-\kappa+\frac{1}{2N(x)^{2}}\left[2E_{1}E_{2}+\kappa(E_{1}^2+E_{2}^2)\right].
\end{equation}

For the case $\kappa=\pm1$, when x goes to zero, the CM energy remains finite
for radially moving particles 
\begin{equation}
\frac{E_{c.m.}^{2}}{2}|_{x=0}\rightarrow 1-\kappa.
\end{equation}

This suggests that although one of the particle is boosted by the naked singularity. There is not any unlimited collisional energy near such a singularity. Note that Compton- like processes were considered first in \cite{piran}, where rotational effect of Kerr black hole played a significant role. Our case here is entirely free of rotational effects.
\section{PHOTON VERSUS AN INFALLING PARTICLE}

A massless photon can naturally scatter an infalling
particle or vice versa. This phenomenon  is analogous to a Compton
scattering taking place in 1+1-dimensions. Null-geodesics for a photon can
be described simply by

\begin{eqnarray}
\frac{dt}{d\lambda } &=&\frac{E_{1}}{N^{2}} \\
\frac{dx}{d\lambda } &=&\pm \sqrt{E_{1}^{2}-N^{2}}  \notag
\end{eqnarray}

where $\lambda $ is an affine parameter and $E_{1}$stands for the photon
energy. Defining $E_{1}=\hslash \omega _{0}$, where $\omega _{0}$ is the
frequency ( with the choice $\hslash =1)$ we can parametrize energy of the
photon \ by $\omega _{0}$ alone. The CM energy of a photon and the infalling particle can be taken now as%
\begin{equation}
E_{cm}^{2}=-(p^{\mu }+k^{\mu })^{2}
\end{equation}

in which $p^{\mu }=mu^{\mu}$ and $k^{\mu }$ refer to the particle and photon, 2-
momenta, respectively. This amounts to%
\begin{equation}
E_{cm}^{2}=m^{2}-2mg_{\mu \nu }u^{\mu }k^{\nu },
\end{equation}

where we have for the particle 
\begin{equation}
p^{\mu }=m\left( \frac{E_{2}}{N^{2}},\sqrt{E_{2}^{2}-N^{2}}\right)
\end{equation}
and for the photon 
\begin{equation}
k^{\mu }=\left( \frac{E_{1}}{N^{2}},-E_{1}\right) .
\end{equation}
One obtains%
\begin{equation}
E_{cm}^{2}=m^{2}+\frac{2mE_{1}}{N^{2}}\left( E_{2}+\kappa \sqrt{%
E_{2}^{2}-N^{2}}\right) .
\end{equation}

In the near horizon limit this reduces to%
\begin{equation}
E_{cm}^{2}=m^{2}+\frac{2mE_{1}}{N^{2}}\left( E_{2}+\kappa E_{2}-\frac{N^{2}}{%
2E_{2}}\right) . \label{61}
\end{equation}

Note that for $\kappa =-1$ we have $\ E_{cm}^{2}$ given by%
\begin{equation}
E_{cm}^{2}=m^{2}\left( 1-\frac{E_{1}}{mE_{2}}\right)
\end{equation}

which is finite between the collision of a photon and an infalling
particle, and therefore is not of interest. As a matter of fact the occurrence of outgoing photon from the event horizon can not be justified unless an explosive/ decay process is assumed to take place. As a result for $\kappa =+1
$ from eq.(\ref{61}) we obtain an unbounded $E_{cm}^{2}$ between the  collision of infalling
photon and particle. Let us add that `inverse' Compton process in the ergosphere of Kerr black hole was considered in \cite{piran} where the photon's energy showed increment due to rotational and curvature effects. The energy, however, attained an upper bound which was finite. Our result obtained here being entirely radial on the other hand can hardly be compared with those of \cite{piran}.

\section{CONCLUSION}

 Our aim was to investigate whether
the BSW type effect which arises in higher dimensional black holes applies
also in the 1+1- D naked singularity/ black hole. The theory we adapted is not general relativity but
instead the recently popular HL gravity. We employed the class of 5-
parametric black hole/ naked singularity solutions found recently \cite{costa}. The class has
particular limits of flat Rindler, Schwarzschild and Reissner-Nordstrom-
like solutions. For each case we have calculated the center-of-mass\ (CM)
energy of the particles and shown that the energy can grow unbounded for
some cases. In other words the strong gravity near the event horizon effects
the collision process with unlimited source to turn it into a natural
accelerator. The model we use applies also to the case of a photon/particle
collision with different characteristics. It is observed that the CM energy
of the infalling particles  from rest at infinity will remain finite in the
CM frame at the event horizon of a black hole. Contrariwise, unlimited CM
energy will be attained between the collision of the outgoing particles from the event horizon region and infalling particles. It is also possible to achieve the infinite energy
between an infalling photon and an infalling massive particle. However, we found  finite CM energy between an outgoing photon and infalling
particle. Finally, we must admit that absence of rotational effects in 1+1-
D restricts the problem to the level of a toy model in which particles move on pure radial geodesics yielding finite CM energy in the vicinity of a naked singularity.

\section{acknowledgement}
We wish to thank the anonymous referees for valuable suggestions. This work was presented as a poster at Karl Schwarzschild
Meeting 20-24 July 2015 Frankfurt Institute for Advanced Studies.

\end{document}